\documentclass[aps,prd,11pt,notitlepage,nofootinbib,superscriptaddress,showkeys,showpacs]{revtex4-2}

\pdfoutput=1

\usepackage{amstext,amsmath,amssymb,amsfonts,bbm,euscript,color,dsfont}

\usepackage[utf8]{inputenc}

\usepackage{epsfig}

\usepackage{hyperref}

\usepackage{slashed}

\usepackage{amsthm}

\usepackage{subfigure}

\usepackage{color}

\usepackage{xcolor}

\usepackage{multirow}

\usepackage{bbold}

\usepackage{tikz}

\usetikzlibrary{arrows,backgrounds}

\usepackage{verbatim} 

\usepackage{graphicx}

\usepackage{latexsym}

\usepackage[shortlabels]{enumitem}

\begin{document}


\title{Physics-informed neural networks viewpoint for solving the Dyson-Schwinger equations of quantum electrodynamics}



\author{Rodrigo Carmo Terin}

\email{rodrigo.carmo@urjc.es}




\affiliation{King Juan Carlos University, Department of Physics, Av. del Alcalde de Móstoles, 28933, Madrid, Spain}





%















%




\begin{abstract}
Physics-informed neural networks (PINNs) are employed to solve the Dyson--Schwinger equations of quantum electrodynamics (QED) in Euclidean space, with a focus on the non-perturbative generation of the fermion's dynamical mass function in the Landau gauge. By inserting the integral equation directly into the loss function, our PINN framework enables a single neural network to learn a continuous and differentiable representation of the mass function over a spectrum of momenta. Also, we benchmark our approach against a traditional numerical algorithm showing the main differences among them. Our novel strategy, which is expected to be extended to other quantum field theories, is the first step towards forefront applications of machine learning in high-level theoretical physics.
\end{abstract}

\maketitle



\section{Introduction and motivation}\label{sec:intro}

The predictions of the behavior of quantum field theories require an in-depth knowledge of the interactions among fields and particles on different energy scales. One of the theoretical methods to study these interactions is through the famous Dyson-Schwinger equations (DSEs), which are known for being an infinite set of integral equations fundamental to investigate the infrared (IR) region in QFTs, particularly in QED \cite{Dyson1949, Schwinger1951,Schwinger1951_b}. These integral equations are responsible for describing the dynamics of n-point Green's functions, which have an importance in investigating phenomena like dynamical mass generation, confinement, and the nature of phase transitions in quantum systems \cite{Roberts1994, Maris1999,Alkofer:2000wg, Huber:2018ned, Roberts:2000aa, Fischer:2018sdj, PhysRevLett.90.152001,Oliveira2024}.

In terms of QED, the fermion and photon propagators' DSEs are fundamental in understanding how gauge invariance and chiral symmetry breaking manifest in different energy regimes \cite{Cornwall1982, Curtis1990, Curtis1991, Marciano1978, Bender2023}. It is important to mention that these integral equations form an infinite tower of equations, where each Green function is connected to higher-order ones, and therefore practical applications in general involve truncating the system. The Rainbow-Ladder approximation, for instance, is a commonly used truncation that simplifies the fermion-boson interaction vertex to its lowest-order term \cite{Atkinson1990, Maris1998, Fischer2004, Rojas2008, Kizilersu2015}, a method often employed to study hadronic physics and the high-energy behavior of the quark-gluon vertex in quantum chromodynamics (QCD) \cite{Roberts1994, Eichmann:2016yit, Ding:2022ows}.

In recent years, machine learning (ML) algorithms have been recognized as important tools for addressing difficult high-dimensional problems in physics. Among these approaches, the PINNs have emerged as particularly promising. These networks integrate the physical laws governing the system directly into the architecture of the neural network, using loss functions informed by the residuals of the differential equations that must be solved \cite{Raissi2019, Karniadakis2021, Lagaris1998}. This method has been effectively applied to solve forward and inverse problems involving non-linear partial differential equations (PDEs), enabling high-precision solutions even in the presence of noisy and incomplete data \cite{Raissi2019, Karniadakis2021}.

The wide-range applicability of PINNs covers some physical areas, e.g., the fluid dynamics, where they have been used to reconstruct flow fields from partial observations \cite{Raissi2020}. In cardiovascular flow modeling, these networks have successfully predicted arterial blood pressure from non-invasive 4D flow Magnetic Resonance Imaging (MRI) data \cite{Kissas2020}, and in plasma physics, they have been used to uncover turbulent transport at the edge of magnetic confinement fusion devices \cite{Mathews2020}. They have also demonstrated utility in quantum chemistry, where they have been designed to handle high-dimensional quantum many-body problems, such as solving the Schr\"{o}dinger equation \cite{Pfau2020, Brevi_2024, technologies12100174}.

Moreover, recent advances have expanded the scope of PINNs; for instance, the introduction of Bayesian physics-informed neural networks (B-PINN) has further improved the ability to quantify uncertainty in predictions, making these networks stronger in scenarios where data may be sparse or noisy \cite{Yang2021b}. In addition, the development of extended PINNs (XPINNs) has enabled more efficient training in parallel architectures \cite{Jagtap2020}.
Other extensions of these networks, such as deep operator networks (DeepONets), were responsible for learning mappings between infinite-dimensional function spaces, allowing the solution of operator learning problems \cite{Lu2021}. There are more examples like multi-fidelity PINNs that combine data from different sources to increase model accuracy and to reduce computational costs \cite{Meng2020}. Furthermore, applications in molecular simulations have led PINNs to accurately predict potential energy surfaces and simulate molecular dynamics \cite{Zhang2018}.

In this work, we extend the use of PINNs to solve the DSEs for the fermion propagator in QED, particularly in Euclidean space. Unlike their conventional applications, which focus on solving PDEs, we employ these networks to tackle the integral equations that govern quantum field theories, such as those represented by DSEs, which are fundamentally more challenging due to their non-local nature and the need to account for all possible interactions in the quantum field~\cite{Fukuda1976, Efimov1970}.

Our adaptation involves incorporating the integral equations of the DSEs directly into the loss function used during training. By doing so, we preserve the full structure and non-local nature of the DSEs. The neural networks approximate the wave function renormalization, the dynamical mass function, and are trained to minimize the discrepancy between their outputs and the solutions to the integral equations. This is achieved by first expressing the momentum variable in dimensionless form, and discretizing a desired interval using a logarithmic grid of $N$ points. Then, the integrals are numerically approximated using the trapezoidal rule ~\cite{BurdenFaires2010,Atkinson1989,SuliMayers2003}, which splits the integration domain into small intervals and approximates the area under the curve by summing the areas of the corresponding trapezoids. In each sub-interval, the quantity of interest is evaluated using our neural network, and these computations are directly incorporated into the loss function.

To confirm numerical stability and physical credibility, we implement positivity of the output through a softplus activation and improve resolution over several momentum scales by adding specific loss terms for intermediate, high, and ultraviolet (UV) regions. This multiscale supervision strategy enables the network to agree with numerical state-of-the-art traditional solutions even when $B(p^2)$ scales many orders of magnitude. In contrast to our previous approach relying only on base residuals, our novel strategy introduces log-scale constraints and perturbative tail matching to capture the full behavior of $B(p^2)$ on IR and UV regimes. As a consequence, the networks learn solutions that are consistent with both the mathematical form of the DSEs and the underlying physical principles they represent, without modifying the network architecture to include integral operators directly.

Finally, our research contributes to a effort to develop physics-informed learning methods capable of solving different problems in distinct scientific domains. By inserting intrinsic physical knowledge within PINNs, this study aims to provide efficient tools for investigating quantum field dynamics. The results could be a significant first step towards novel applications of machine learning in physics, from high-energy particle collisions to condensed matter systems.

This paper is organized as follows. In Sec.~\ref{sec:background}, we briefly review the Dyson--Schwinger equations in Euclidean space, presenting the fundamental expressions for the fermion and photon propagators. Then, in Sec.~\ref{sec:NN_approach}, we introduce the rainbow truncation framework and detail our neural network solution strategy, including the trapezoidal numerical integration method, the network architectures, and the training procedure (hyperparameters, loss function, etc.). Next, in Sec.~\ref{sec:results}, we present and analyze the results attained from our PINNs, focusing on the wave function renormalization and the dynamical mass function. Also, we compare our framework against a traditional numerical algorithm. Lastly, Sec.~\ref{sec:conclusion} concludes the manuscript, summarizing our findings and suggesting potential directions for future research.


\section{Background of the Dyson-Schwinger equations in Euclidean space}\label{sec:background}

The Dyson-Schwinger equations provide a non-perturbative framework for determining Green’s functions in quantum field theories. In this work, we focus on the DSEs for the renormalized fermion in QED within Euclidean space by following the theoretical formalism developed in our previous work~\cite{Oliveira:2022bar}. The transition from Minkowski space-time to Euclidean space is performed through a Wick rotation, which converts the Minkowski time component $p_0$ into an imaginary component $p_4 = i p_0$. This results in the Euclidean metric $p^2 = p_1^2 + p_2^2 + p_3^2 + p_4^2$, simplifying the analytic structure of the integrals.

In Euclidean space, the inverse renormalized fermion propagator $S_E^{-1}(p)$ can be written in terms of the wavefunction renormalization $A(p^2)$ and the mass function $B(p^2)$:
\begin{equation}
S_E^{-1}(p) = i \slashed{p}A(p^2)  + B(p^2),
\label{eq:inverse_fermion_prop}
\end{equation}
where $\slashed{p} = \gamma_\mu p_\mu$ and the Euclidean gamma matrices $\gamma_\mu$ satisfy the anticommutation relation $\{ \gamma_\mu, \gamma_\nu \} = 2 \delta_{\mu\nu}$. Furthermore, we also have the ratio $M(p^2) = \frac{B(p^2)}{A(p^2)}$ that defines the momentum-dependent mass function of the fermion. It characterizes the non-perturbative dynamical generation of a momentum-dependent mass for the fermion, even if the bare mass $m$ is zero. In general, both $A(p^2)$ and $B(p^2)$ are determined by solving the coupled system of integral equations derived from the DSEs.

To make the DSEs tractable, we use the rainbow approximation~\cite{Roberts1994,Maris1998}, which truncates the infinite set of coupled integral equations down to a manageable form while retaining essential non-perturbative features. The bare fermion-photon vertex is given by:
\begin{equation}
\Gamma^\mu(p, k) = \gamma^\mu.
\label{rain}
\end{equation}

The functions $A(p^2)$ and $B(p^2)$ satisfy the following coupled integral equations derived from the DSEs:
\begin{align}
B(p^2) &= m_{ph} + g_{ph}^2 \bigl[ \Sigma_s(p^2) - \Sigma_s(\mu_F^2) \bigr], \label{eq:Bp2}\\[6pt]
A(p^2) &= 1 - g_{ph}^2 \bigl[ \Sigma_v(p^2) - \Sigma_v(\mu_F^2) \bigr], \label{eq:Ap2}
\end{align}
Here, $\Sigma_s$ and $\Sigma_v$ are the scalar and vector components of the self-energy of the fermion. Moreover, $m_{ph}$ is the renormalized physical mass of the fermion, and the coupling is defined as $g_{ph}^2 = 4\pi\alpha$, where $\alpha$ is the effective fine-structure constant in the theory. These quantities are defined at the renormalization scale $\mu_F^2$, and their relation to the bare parameters is established through the renormalization conditions discussed later in this section.

The fermion self-energy in Euclidean space is expressed as:
\begin{equation}
\Sigma(p) = - g_{ph}^2 \int \frac{d^4 k}{(2\pi)^4} \gamma_\mu S(p - k) \gamma_\nu D_{\mu\nu}(k).
\end{equation}
The fermion propagator is:
\begin{equation}
S(p - k) = \frac{-A((p - k)^2) \bigl(i(\slashed{p} - \slashed{k})\bigr) + B((p - k)^2)}{A^2((p - k)^2) (p - k)^2 + B^2((p - k)^2)}\,,
\end{equation}
while the photon propagator is decomposed as:
\begin{equation}
D_{\mu\nu}(k) = \left(\delta_{\mu\nu} - \frac{k_\mu k_\nu}{k^2}\right) D(k^2) + \xi \frac{k_\mu k_\nu}{k^4},
\end{equation}
which $\xi$ is the gauge parameter. Using the gamma matrix identities:
\begin{equation}
\gamma_\mu \gamma_\nu = \delta_{\mu\nu} - i \sigma_{\mu\nu}, \quad \{\gamma_\mu, \gamma_\nu\} = 2 \delta_{\mu\nu},
\end{equation}
where $\sigma_{\mu\nu} = \frac{i}{2}[\gamma_\mu, \gamma_\nu]$, we can now decompose the fermion self-energy $\Sigma(p)$ into its scalar and vector components. Since $\Sigma_s(p^2)$ and $\Sigma_v(p^2)$ are defined without including the coupling factor, we explicitly factor out $g_{ph}^2$:
\begin{equation}
\Sigma(p) = g_{ph}^2 \bigl[ (i \slashed{p})\,\Sigma_v(p^2) + \Sigma_s(p^2) \bigr].
\end{equation}

To determine $\Sigma_s(p^2)$ and $\Sigma_v(p^2)$, one performs the gamma matrix algebra, applies the integrals to the loop momentum, and identifies the coefficients of the scalar and vector structures. The scalar component $\Sigma_s(p^2)$ is attained from terms that do not involve gamma matrices (after all contractions), whereas $\Sigma_v(p^2)$ is extracted from terms proportional to $i \slashed{p}$.

Although the integral expressions for $\Sigma_s$ and $\Sigma_v$ are formally similar, the vector part $\Sigma_v(p^2)$ does not simplify easily due to the non-trivial momentum dependence of $A, B,$ and $D$. A representative form of these integrals is:
\begin{align}
\Sigma_s(p^2) &= \int \frac{d^4 k}{(2\pi)^4} \frac{\bigg[3 D(k^2) B((p - k)^2) + \xi\, \frac{B((p - k)^2)}{k^2}\bigg]}{A^2((p - k)^2)(p - k)^2 + B^2((p - k)^2)}, \\[6pt]
\Sigma_v(p^2) &= \int \frac{d^4 k}{(2\pi)^4} \frac{A((p - k)^2)(p - k)^2\bigg[3D(k^2) + \xi\, \frac{1}{k^2}\bigg]}{\bigg[A^2((p - k)^2)(p - k)^2 + B^2((p - k)^2)\bigg] p^2}.
\end{align}

The photon propagator satisfies its own DSE:
\begin{equation}
\frac{1}{D(k^2)} = k^2\bigl[1 - g_{ph}^2(\Pi(k^2)-\Pi(\mu_B^2))\bigr],
\end{equation}
where the photon self-energy $\Pi(k^2)$ is given by:
\begin{equation}
\Pi(k^2) = \int \frac{d^4 p}{(2\pi)^4} F(p^2) F((p - k)^2) \bigl[A(p^2) A((p - k)^2)(p^2 - p\cdot k) + 2 B(p^2) B((p - k)^2)\bigr],
\end{equation}
with
\begin{equation}
F(p^2) = \frac{1}{A^2(p^2) p^2 + B^2(p^2)}.
\end{equation}

Renormalization in QED involves distinct renormalization constants: $Z_2$ for the fermion field, $Z_m$ for the mass, and $Z_A$ for the photon field. The physical mass and coupling at the scale $\mu_F$ relate to the bare parameters as:
\begin{equation}
m_{ph} = Z_2 m - g^2 \Sigma_s(\mu_F^2),
\end{equation}
where \(Z_2 = \frac{1}{1 - g^2 \Sigma_v(\mu_F^2)}\). The renormalized coupling is defined through:
\begin{equation}
g_{ph} = \frac{g}{Z_g}, \quad \text{with } Z_g = \frac{1}{\sqrt{Z_A}}.
\end{equation}

By specifying the appropriate renormalization conditions in $\mu_F$, one fixes $Z_2, Z_m, Z_A$, and thus determines the physical parameters $m_{ph}$ and $g_{ph}$. In this way, all renormalized quantities are manifested finite and well defined at the chosen renormalization scale.

\section{The neural network approach}
\label{sec:NN_approach}

Our main focus in this work is to approximate the dynamical mass function of the fermion through a single PINN in the Landau gauge, where the gauge parameter $\xi$ is equal null and compare our results with a traditional (conservative) numerical algorithm and with works \cite{Kizilersu:2014ela, Williams:2007zzh}. Under these conditions, the wave function renormalization is effectively $1$ at leading order, which simplifies the DSEs considerably. Consequently, the mass function is given directly by
\begin{equation}
M(p^2) = B(p^2).
\label{mass}
\end{equation}
While more general scenarios could involve PINNs for $A(p^2)$, $B(p^2)$, and $D(k^2)$ simultaneously, our present effort focuses on learning $B(p^2)$, since the photon propagator is ill-defined in the above-mentioned gauge by using the DSEs formalism \cite{Oliveira:2022bar}. To do so we start from the full DSE as given in Eq.~\eqref{eq:Bp2}, we then apply a series of approximations to derive its simplified form. First, we employ the rainbow approximation by replacing the dressed fermion–photon vertex with its bare form Eq.~\eqref{rain}, and approximating the photon propagator by its leading-order expression. Although such simplifications are standard in the study of dynamical mass generation, we note that the structure of our PINN architecture is, in principle, flexible enough to incorporate more sophisticated truncations, including gauge-independent quantities, as long as  the corresponding integral expressions are properly implemented in the loss function. This potential extension is currently under investigation \cite{Terin:InPrep}. We then choose renormalization conditions that remove the subtraction terms, or equivalently, we work in the chiral limit where $m_{ph}=0$; as a consequence, the scalar part simplifies to $B(p^2) \approx g_{ph}^2\,\Sigma_s(p^2)$. Following this, the angular integrations in four-dimensional Euclidean space are performed, which reduce the full loop integral to a one-dimensional integral over $k^2$ with an integration kernel that naturally splits into two regions. Specifically, for $0\le k^2 \le p^2$ the kernel is $K(p^2,k^2)=\dfrac{k^2}{p^2}$, while for $p^2 \le k^2 \le \kappa^2$ it becomes $K(p^2,k^2)=1$. By incorporating these results, the scalar self-energy can be expressed as
\[
\Sigma_s(p^2) \propto \int_0^{\kappa^2} \frac{B(k^2)}{k^2+B^2(k^2)}\,K(p^2,k^2)\,dk^2.
\]
Finally, combining all these approximations leads to the simplified integral equation for \(B(p^2)\):
\begin{equation}
B(p^2)
\;=\;
\frac{3\,\alpha}{4\pi}
\Biggl[
\int_{0}^{p^2} \frac{B(k^2)}{k^2 + B^2(k^2)} \,\frac{k^2}{p^2}\,dk^2
\;+\;
\int_{p^2}^{\kappa^2} \frac{B(k^2)}{k^2 + B^2(k^2)} \,dk^2
\Biggr],
\label{eq:B_landau_equation}
\end{equation}
where $\kappa^2$ is a fixed infrared scale (often set to $1$ for convenience). In what follows, we outline our neural network structure, describe how we numerically approximate the integrals, and detail the training procedure that implements Eq.~\eqref{eq:B_landau_equation}.


\subsection{Neural network architecture}

We apply a single neural network $model\_B$ that approximates $B(p^2)$. This network has:
\begin{enumerate}
    \item An input layer of dimension $1$ receiving \(x = p^2/\kappa^2\) on a logarithmic grid (e.g., \(x \in [10^{-12},\,1]\)).
    \item Three hidden layers, each with $64$ neurons, using the hyperbolic tangent ($\tanh$) activation function.
    \item An output layer of dimension $1$ with a softplus activation to ensure $B_{\mathrm{pred}}(x) > 0$. That is, the predicted value is given by $B_{\mathrm{pred}}(x) = \log(1 + \exp(\tilde{B}(x)))$.
\end{enumerate}

This architecture better approximate $B(p^2)$ over several orders of magnitude, especially in the UV region where the values become very small. The positivity constraint is automatically implemented by the softplus function, which also avoids potential numerical instabilities in the integral denominators $k^2 + B^2(k^2)$.

The network parameters (weights and biases) are initialized by default with Glorot uniform initialization~\cite{Glorot2010}, and the bias in the final layer can optionally be initialized to $\log(\exp(10^{-3}) - 1)$ to make $B \approx 10^{-3}$ at the start of training.


\subsection{Numerical approximation of the integral}

Since $A(p^2)=1$ in Landau gauge at leading order, the self-energy expression simplifies to Eq.~\eqref{eq:B_landau_equation}. We rewrite $p^2$ in dimensionless form $x = p^2/\kappa^2$, and similarly $k^2/\kappa^2$ for the integration variable. Thus, the integrals become
\begin{equation}
B(x)
\;=\;
\frac{3\,\alpha}{4\pi}\biggl[
\underbrace{\int_{0}^{x}\! \frac{B(k)}{k + B^2(k)}\;\frac{k}{x}\;dk}_{\text{IR part}}
\;+\;
\underbrace{\int_{x}^{1}\! \frac{B(k)}{k + B^2(k)}\;dk}_{\text{UV part}}
\biggr].
\label{eq:integral_B_x}
\end{equation}
Here, \(0\le x \le1\). In practice, we discretize $[10^{-12},\,1]$ using a logarithmic grid of $N$ points, $\{x_1, x_2,\dots,x_N\}$, and approximate the integrals using a trapezoidal rule. For the interval $[0,x_i]$, we sum over $\{x_j\}_{j=1}^{i}$; for $[x_i,1]$, we sum over $\{x_j\}_{j=i}^{N}$. Symbolically,
\begin{align}
\int_{0}^{x_i}\!f(k)\,dk &\;\approx\; \sum_{j=1}^{i-1}\! \tfrac12\,\bigl[f(x_j) + f(x_{j+1})\bigr]\;\bigl[x_{j+1}-x_j\bigr],\\[6pt]
\int_{x_i}^{1}\!g(k)\,dk &\;\approx\; \sum_{j=i}^{N-1}\! \tfrac12\,\bigl[g(x_j) + g(x_{j+1})\bigr]\;\bigl[x_{j+1}-x_j\bigr].
\end{align}

In each trapezoid sub-interval, we evaluate $B(x_j)$ using our neural network. The coupling $\alpha$ is written as a constant in front of the integral, but it can also be treated as a parameter to be fitted.

To improve precision in specific momentum regions, we define masks over subsets of the domain: a “mid" region ($10^{-7} \le x \le 10^{-3}$), “high" ($10^{-3} \le x \le 10^{-1}$), and “very high" ($10^{-1} \le x \le 1$). These intervals will be used in the loss function to guide the network with which we refer to as regional supervision, this strategy consists of incorporating localized loss contributions into the total loss function, enabling the network to balance its performance over multiple orders of magnitude.

Moreover, a logarithmic comparison between the network prediction and the traditional solution is included in each region to better work out the dynamic range of $B(x)$. This will be fundamental to ensure that the PINN matches the benchmark results both qualitatively and quantitatively on all energy scales.

\subsection{Training procedure and loss construction}

Let us show now our PINN strategy, i.e., at both each epoch and discrete point $x_i$, we compute:
\[
B_{\mathrm{pred}}(x_i)
\;=\;
\texttt{model\_B}(x_i),
\]
and also form
\[
B_{\mathrm{target}}(x_i) \;=\;
\frac{3\,\alpha}{4\pi}
\Biggl[
\int_{0}^{x_i}
\frac{B(k)}{k + B^{2}(k)} \,\frac{k}{x_i}\,\mathrm{d}k
\;+\;
\int_{x_i}^{1}
\frac{B(k)}{k + B^{2}(k)}\,\mathrm{d}k
\Biggr]\,,
\]
through the trapezoidal sums. We then store up a mean-squared error (MSE) over all points $x_i$:
\begin{equation}
\text{Loss}_{\mathrm{base}} \;=\; \frac{1}{N} \sum_{i=1}^{N}
\Bigl[
B_{\mathrm{pred}}(x_i) \;-\; B_{\mathrm{target}}(x_i)
\Bigr]^2.
\end{equation}

To improve learning over several orders of magnitude of $B(p^2)$, especially for values $\lesssim 10^{-6}$, additional loss terms were introduced. These include:

\begin{enumerate}[label=\roman*.]
    \item A mid-momentum range loss:
    \begin{equation}
    \text{Loss}_{\mathrm{mid}} = \frac{1}{N_{\mathrm{mid}}} \sum_{x_i \in [10^{-7},\,10^{-3}]} 
    \left[\log_{10} B_{\mathrm{pred}}(x_i) - \log_{10} B_{\mathrm{num}}(x_i) \right]^2\,.
    \end{equation}

    \item A high-momentum loss in the interval $[10^{-3},\,10^{-1}]$:
    \begin{equation}
    \text{Loss}_{\mathrm{high}} = \frac{1}{N_{\mathrm{high}}} \sum_{x_i \in [10^{-3},\,10^{-1}]} 
    \left[\log_{10} B_{\mathrm{pred}}(x_i) - \log_{10} B_{\mathrm{num}}(x_i) \right]^2\,.
    \end{equation}

    \item A perturbative UV tail constraint:
    \begin{equation}
    \text{Loss}_{\mathrm{tail}} = \frac{1}{N_{\mathrm{uv}}} \sum_{x_i \in [10^{-2},\,1]} 
    \left[x_i B_{\mathrm{pred}}(x_i) - C \right]^2,
    \end{equation}
    where $C$ is the mean $C = \langle x_i B_{\mathrm{num}}(x_i) \rangle$ in the UV.
\end{enumerate}

The final loss is a weighted sum of all components:
\begin{equation}
\text{Loss}_{\mathrm{total}} = \text{Loss}_{\mathrm{base}} + 
\lambda_{\mathrm{mid}} \text{Loss}_{\mathrm{mid}} +
\lambda_{\mathrm{high}} \text{Loss}_{\mathrm{high}} +
\lambda_{\mathrm{tail}} \text{Loss}_{\mathrm{tail}},
\end{equation}
with chosen weights $\lambda_{\mathrm{mid}} = \lambda_{\mathrm{high}} = 10^6$, and $\lambda_{\mathrm{tail}} = 10^2$. Moreover, $N_{\mathrm{mid}}, N_{\mathrm{high}},$ and $N_{\mathrm{uv}}$ denote the number of grid points $x_i$ lying in the intervals $[10^{-7},\,10^{-3}]$, $[10^{-3},\,10^{-1}]$, and $[10^{-2},\,1]$, respectively. These quantities correspond to the cardinality of the masked subsets used during loss computation, and serve to normalize the loss terms on different momentum regions, ensuring that no interval dominates due to a larger number of points. Also, this composite loss structure enables the network to reproduce the state-of-the-art traditional (conservative) numerical results. To accomplish that, we use the automatic differentiation capabilities of JAX~\cite{jax2018github} together with the neural network infrastructure of Equinox~\cite{kidger2021equinox} and the optimization routines provided by Optax~\cite{deepmind2020optax}. These libraries let efficient gradient-based optimization of the neural parameters. Furthermore it is worth mentioning that our learning strategy is not a standard supervised learning approach. Although some components of the loss function involve comparisons with results from the traditional numerical algorithm, these are used only as auxiliary regularization terms to stabilize training in specific momentum regions. The core of our method lies in minimizing the residual of the DSEs on the input domain, following the PINN paradigm. In this sense, the training falls under the category of residual-based physics-informed learning, rather than supervised or unsupervised learning in the classical machine learning taxonomy.

\section{Discussion of results} \label{sec:results}

In this section, we present our solutions attained from both a traditional numerical iterative solver and our PINN approach for $A(p^2)$ and $B(p^2)$ in Landau gauge.


\subsection{Traditional and PINNs methods}

First, we have implemented a traditional (conservative) numerical method to attain solutions of the DSEs in the above mentioned gauge, serving as a benchmark for our PINN-based approach. Using as references the works \cite{Kizilersu:2014ela, Williams:2007zzh}, our simulations show how variations in the coupling (instead of an explicit UV cutoff parameter) can affect the non-perturbative features of the theory. In this iterative method, the equation
\begin{equation}
B(p^2)
\;=\;
m + \frac{3\,\alpha}{4\pi}\left[
\int_{0}^{p^2} \frac{B(k^2)}{k^2+B^2(k^2)} \,\frac{k^2}{p^2}\,dk^2
+\int_{p^2}^{\kappa^2} \frac{B(k^2)}{k^2+B^2(k^2)}\,dk^2
\right]
\label{eq:iterative_B}
\end{equation}
is discretized over a logarithmic grid, with $x = p^2/\kappa^2$ ranging over the interval $[10^{-12},\,1]$. The iterative scheme initializes the function with $B=10^{-3}$, and then updates it using the trapezoidal rule until convergence (with tolerance $10^{-7}$). In parallel, we train our PINN using the same input domain and reference values for the coupling $\alpha$, while introducing a loss function with additional terms targeting specific momentum intervals as written before.

\begin{figure}[ht]
    \centering
    \includegraphics[width=0.9\textwidth]{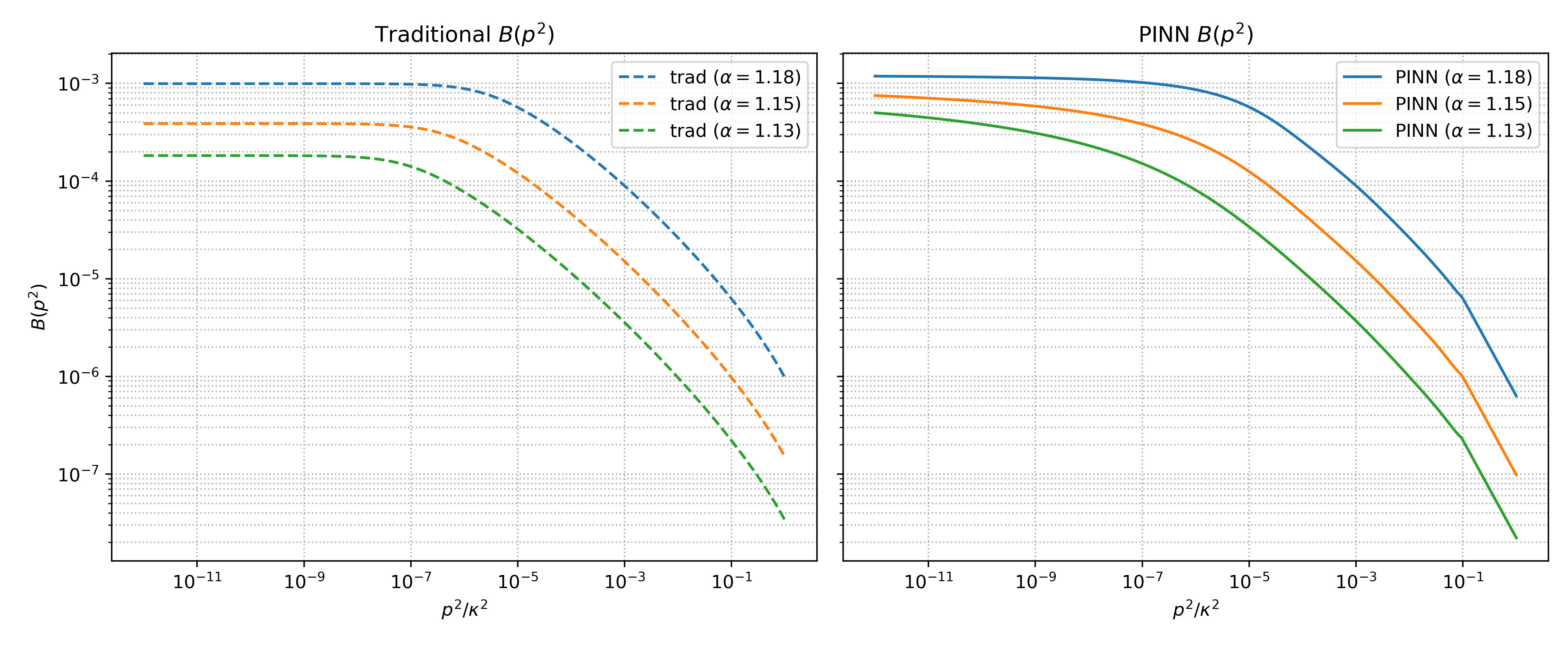}
    \caption{Comparison between traditional numerical solution (dashed) and PINN results (solid) for the fermion dynamical mass function $B(p^2)$ at $\alpha = 1.13,\ 1.15,\ 1.18$.}
    \label{fig:B_joint}
\end{figure}

Figure~\ref{fig:B_joint} displays the results for the mass function $B(p^2)$ obtained from both approaches for three values of the coupling $\alpha = 1.13,\ 1.15,\ $ and $1.18$, over the interval $p^2/\kappa^2 \in [10^{-12},\,1]$. The traditional results are shown with dashed lines, while the corresponding PINN predictions are shown with solid lines of matching color. This combined plot shows that our PINN reproduces the non-perturbative behavior on all momentum scales, from deep IR to UV, including features down to $B\sim 10^{-7}$ in the UV.

We also observe that the agreement is especially accurate in the intermediate and UV regimes, where our earlier PINN implementation plateaued. This is a direct consequence of introducing log-scale losses and regional supervision in the training.

\begin{figure}[ht]
    \centering
    \includegraphics[width=0.9\textwidth]{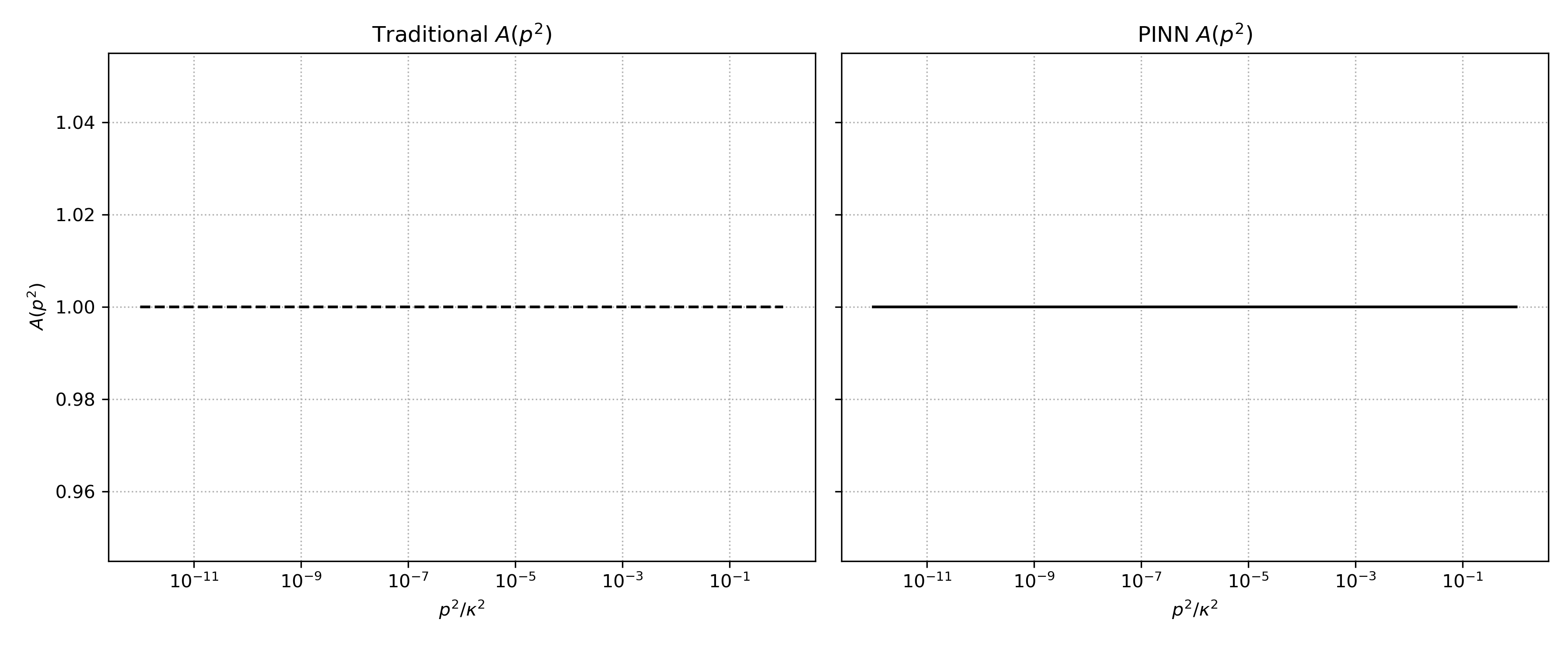}
    \caption{Fermion wave function renormalization $A(p^2)$, fixed to $1$ in Landau gauge.}
    \label{fig:A_landau}
\end{figure}

As expected, increasing the coupling $\alpha$ increases the non-perturbative effects in the infrared region, yielding a larger value of $B(p^2)$ at low $p^2$. For all cases, the wave function renormalization remains fixed at $A(p^2)=1$ in Landau gauge \cite{Kizilersu:2014ela, Williams:2007zzh}, shown in Fig.~\ref{fig:A_landau}, and used consistently in both methods.

\newpage
We emphasize that the results shown in Figs.~\ref{fig:B_joint} and ~\ref{fig:A_landau} were attained using the updated PINN framework presented in this work, which includes explicit loss weighting on logarithmic scales and targeted supervision in the mid, high, and very high momentum intervals. These additions were fundamental to suppress the PINN’s tendency to plateau and to ensure alignment with the expected numerical behavior.

It is worth commenting that the references \cite{Kizilersu:2014ela, Williams:2007zzh} focus on the so-called rainbow approximation of the fermion DSE in $4$D quantum electrodynamics, in which the photon propagator is kept quenched and the vertex is taken at its bare form $\gamma^{\mu}$. Under this approximation, one investigates the conditions under which dynamical mass generation emerges, typically identifying a critical coupling $\alpha_c$ above which $M(p^2)$ becomes nonzero. Additionally, the ultraviolet cutoff is implicitly fixed in those analyses, and is not treated as a free parameter; rather, the central interest lies in the gauge parameter $\xi$ and whether $\alpha>\alpha_c$ leads to a spontaneously generated mass (obeying Miransky scaling near $\alpha_c$).

Therefore, in order to adapt our forefront method to the traditional DSE literature, our current perspective is to vary the coupling $\alpha$ directly in the integral equation, thus tracking how the infrared behavior of $B(p^2)$ evolves with different $\alpha$ values. Our viewpoint lines up with the idea of exploring dynamical symmetry breaking through a machine learning approach, while remaining consistent with earlier numerical results of the rainbow approximation.

The agreement over all regions, i.e., IR, intermediate, and UV shows that PINNs, when properly guided with physical constraints and loss shaping, can match traditional numerical methods even for non-local integral equations. This is the first step towards the confirmation of the viability of PINNs as a robust method to tackle DSEs. Finally, although the figures do not explicitly incorporate a variable ultraviolet cutoff, one could indeed combine this framework with a regulator function $R(k^2)$ to examine the sensitivity of the solutions to changing $\Lambda$, which was done in our very first version of this work. 

\subsection{Overwiew by comparing the different perspectives}

As it is already well-known we employed two distinct numerical viewpoints for solving the DSE for the dynamical mass function $B(p^2)$ in Landau gauge. The summary of both of them is given in tables \eqref{tab:traditional} and \eqref{tab:pinn} below

\begin{table*}[h!]
\scriptsize
\centering
\begin{tabular}{|l|p{12cm}|} \hline
\textbf{Characteristic} & \textbf{Traditional numerical method} \\ \hline
Representation & Discrete values of $B(p^2)$ stored on a logarithmic grid from $10^{-12}$ to $1$. \\ \hline
Architecture/Initialization & Started with a small constant for $B(p^2)$ on the grid. \\ \hline
Update mechanism & Iterative, point-by-point updates using trapezoidal integration until convergence. \\ \hline
Convergence criteria & Iterations stop when the maximum difference between successive updates is below $10^{-7}$ or after a maximum number of iterations. \\ \hline
Computational cost & Can become computationally expensive with a refined momentum grid, requiring careful tuning of initial guesses and iteration parameters. \\ \hline
Output & Produces a discrete solution defined only at the grid points. \\ \hline
\end{tabular}
\caption{Characteristics of the traditional numerical method.}
\label{tab:traditional}
\end{table*}

\begin{table*}[h!]
\scriptsize
\centering
\begin{tabular}{|l|p{12cm}|} \hline
\textbf{Characteristic} & \textbf{PINN framework} \\ \hline
Representation & Continuous function $B(x)$ with $x=p^2/\kappa^2$ represented by a neural network model. \\ \hline
Architecture/Initialization & Three hidden layers with $64$ neurons each (using $\tanh$ activation), and a positive output required through softplus activation. \\ \hline
Update mechanism & Global gradient descent (Adam) minimizes an MSE loss that compares $B_{\text{pred}}(x_i)$ with a target $B_{\text{target}}(x_i)$ obtained through numerical integration. \\ \hline
Convergence criteria & Training stops when the optimizer stabilizes or a preset epoch limit is reached. \\ \hline
Computational cost & Loss function includes interval-based supervision and logarithmic penalties, enabling multi-scale resolution at all $p^2$ regions. \\ \hline
Output & Yields a smooth, continuous approximation of $B(p^2)$ valid over the entire domain. \\ \hline
\end{tabular}
\caption{Features of our PINN framework.}
\label{tab:pinn}
\end{table*}


\section{Conclusion} \label{sec:conclusion}

In this work, we have used PINNs to tackle the DSEs for QED defined within the bounds of Euclidean space. In summary, we have shown how a single PINN can be applied to the integral equation for the dynamical mass function in Landau gauge. By discretizing the interval $[10^{-12},1]$ and applying a trapezoidal approximation, we trained a neural network to match $B(p^2)$ against its own integral expression. Our results confirm that this approach is capable of capturing the main non-perturbative features, namely dynamical mass generation at small momenta and a decreasing behavior for larger $p^2$.

Here, we introduced targeted loss components in specific momentum intervals, e.g., mid, high, and ultraviolet regimes combined with logarithmic constraints and perturbative tail matching. This multi-scale approach guarantees that the predicted mass function $B(p^2)$ agrees with traditional (conservative) numerical methods over the entire domain, even down to values $\sim 10^{-7}$. These refinements close the previous performance gap and allow the PINN to serve both as qualitative exploratory tool and as a quantitatively accurate method.

This work represents a first step towards applying modern neural architectures to faithfully reproduce traditional numerical results in non-perturbative QFTs, while retaining differentiability and flexibility for further generalizations. Future research may extend these algorithms within the quantum theoretical physics viewpoint to investigate the same problem in Minkowski space-time, explore more general truncation schemes and other gauges to study gauge (in)dependence explicitly, and employ our approach to different coupled systems, e.g., QCD or scalar Yukawa models, as well as to investigate other $n$-point Green functions. From a computer science perspective other neural network algorithms such as the neuroevolutionary framework \cite{stanley2019designing, Terin:2024kzc}, deep operator networks \cite{Lu2021}, and transformer-based architectures \cite{vaswani2017attention} could share the same ultimate goal of incorporating physical structure into machine learning models to approximate non-local integral equations like DSEs.

Finally, our work intends to extend the use of PINNs to solve difficult Dyson-Schwinger integral equations in QFTs. This approach addresses the missing links between machine learning and theoretical physics and offers a new window into the study of quantum field effects and, we hope, also contributes to the larger efforts to develop computational methods in the era of problems relevant to modern physics.


\section*{Acknowledgments}

R. C. Terin thanks the useful suggestions and constructive comments from the SciPost Physics reviewers. The author is currently a lecturer in Physics at King Juan Carlos University (Madrid, Spain).


\bibliographystyle{unsrt}
\bibliography{DSE}

\end{document}